\documentclass[twocolumn,conference,letterpaper]{IEEEtran}
\usepackage[left=0.75in,right=0.75in,top=1in,bottom=0.75in]{geometry}

\usepackage{cite}
\usepackage{amssymb,amsmath,latexsym,amsfonts,amsthm,mathtools}

\usepackage{graphicx}
\usepackage{float,subcaption}
\usepackage{textcomp}
\usepackage{xcolor}
\definecolor{darkpastelpurple}{rgb}{0.59, 0.44, 0.84}
\usepackage{hyperref}
\hypersetup{colorlinks, breaklinks, citecolor=blue, linkcolor=blue, urlcolor=blue}
\usepackage[nameinlink]{cleveref}
\Crefformat{figure}{#2Fig.~#1#3}
\Crefmultiformat{figure}{Figs.~#2#1#3}{ and~#2#1#3}{, #2#1#3}{ and~#2#1#3}

\usepackage{tikz}
\usetikzlibrary{automata, shapes, arrows, calc, arrows.meta, fit, positioning}

\theoremstyle{plain}
\newtheorem{theorem}{Theorem}

\newtheorem*{problem*}{Problem}
\theoremstyle{remark}
\newtheorem{remark}{Remark}

\newtheorem{definition}{Definition}
\theoremstyle{definition}

\usepackage{mathrsfs}

\usepackage{newtxmath}
\usepackage{booktabs}
\usepackage{duckuments}

\IEEEoverridecommandlockouts

\begin{document}

\title{Safety-Critical Input-Constrained Nonlinear Intercept Guidance in Multiple Engagement Zones}

\author{Praveen Kumar Ranjan,~\IEEEmembership{Member,~IEEE}, Abhinav Sinha,~\IEEEmembership{Senior Member,~IEEE},\\ and Yongcan Cao,~\IEEEmembership{Senior Member,~IEEE}
		\thanks{P. K. Ranjan and Y. Cao are with the Unmanned Systems Lab, Department of Electrical and Computer Engineering, The University of Texas at San Antonio, San Antonio, TX 78249, USA. (e-mails: praveen.ranjan@my.utsa.edu, yongcan.cao@utsa.edu). A. Sinha is with the Guidance, Autonomy, Learning, and Control for Intelligent Systems (GALACxIS) Lab, Department of Aerospace Engineering and Engineering Mechanics, University of Cincinnati, OH 45221, USA (e-mail:abhinav.sinha@uc.edu).}
        }
	
	\maketitle
	\thispagestyle{empty}

    \begin{abstract}
This paper presents an input-constrained nonlinear guidance law to address the problem of intercepting a stationary target in contested environments with multiple defending agents. Contrary to prior approaches that rely on explicit knowledge of defender strategies or utilize conservative safety conditions based on a defender's range, our work characterizes defender threats geometrically through engagement zones that delineate inevitable-interception regions. Outside these engagement zones, the interceptor remains invulnerable. The proposed guidance law switches between a repulsive safety maneuver near these zones and a pursuit maneuver outside their influence. To deal with multiple engagement zones, we employ a smooth minimum function (log-sum-exponent approximation) that aggregates threats from all the zones while prioritizing the most critical threats. Input saturation is modeled and embedded in the non-holonomic vehicle dynamics so the controller respects actuator limits while maintaining stability. Numerical simulations with several defenders demonstrate the proposed method’s ability to avoid engagement zones and achieve interception across diverse initial conditions.
\end{abstract}


\section{Introduction}
The growing need for autonomous systems operating in contested environments necessitates guidance strategies that ensure both survivability and mission accomplishment \cite{doi:10.2514/1.37030,9172808,doi:10.2514/1.51765,doi:10.2514/1.G003157,10634571,10660561,10839025}. Applications such as target interception, air defense, and precision strike missions place requirements on guidance strategies in addition to ensuring negligible miss distance. Such requirements have been addressed in the form of terminal constraints, e.g., impact time (see \cite{9000526,doi:10.2514/1.G005367,doi:10.2514/1.G005180} and references therein). However, in many settings, the pursuer must often breach defended regions to engage with the target and function under strict sensing and maneuvering limitations. A fundamental abstraction for analyzing these interactions is the Target–Attacker–Defender (TAD) engagement, where a pursuer (or attacker) attempts to reach a target while one or more defenders employ interception strategies to neutralize the attacker \cite{sinha2022three,9274339}.

While defender–target cooperation \cite{7171913,9274339,Casbeer2018,sinha2022three,doi:10.2514/6.2010-7876,doi:10.2514/1.58566} has been the primary focus in most prior studies of TAD games, only limited attention has been given to strategies from the attacker’s perspective. In \cite{doi:10.2514/6.2025-1902}, the authors presented a pursuit strategy under an adversarial environment by leveraging reinforcement learning and data-driven methodologies. The three-agent pursuit-evasion dynamics is further investigated in \cite{doi:10.2514/1.61832}, where differential game formulations are employed to derive sufficient conditions for the attacker to strike the target while avoiding interception by the defenders. The authors in \cite{SUN20192337} designed guidance laws to steer the attacker to a defender-safe boundary and then maintain a critical miss distance, establishing attacker-win conditions under linearized game dynamics without requiring knowledge of the target/defender control efforts. In \cite{doi:10.2514/1.51611}, a linear quadratic differential game was used to derive cooperative pursuit–evasion strategies, including the homing interceptor’s optimal pursuit and evasion policy. The authors in \cite{QI20171958} utilized a three-player bounded-control game to design an attacker strategy that guarantees a miss distance from the defender by avoiding the infeasible zero effort miss region and establishes sufficient conditions for the attacker's success. The impact of bounded control effort on the defender's capability was studied in \cite{RUSNAK20119349}.

It is worth mentioning that most of the above-mentioned works require complete knowledge of the defender's strategy to guarantee the attackers' escape and employ simplified vehicle models (e.g., linearized dynamics). This assumption may be impractical in realistic scenarios, especially those involving multiple defenders with range and maneuverability constraints. A more practical approach is to design the intercept guidance strategy using geometric threat sets rather than exact defender threat predictions. An engagement zone (EZ) defines the set of attacker-relative states from which a defender can guarantee interception, assuming the attacker does not change course. Several studies have explored modeling engagement or capture zones under different structural constraints (e.g., static obstacles \cite{OYLER20161}, constrained environments \cite{ZHOU201664}, and visibility \cite{IBRAGIMOV1998187}). An alternate relevant formulation involves modeling the defenders as range-limited, where each defender can travel only up to a maximum range, as considered in \cite{10365808, doi:10.2514/1.I011394,doi:10.2514/1.I011593}. These modeling approaches offer ways to encode threat geometry, which help develop safety-aware guidance strategies in constrained environments. Unlike static safety margins or fixed threat envelopes, EZs are dynamic and geometry-dependent, incorporating key factors such as velocity ratios, turning constraints, and capture radii. 

Despite substantial progress in characterizing the EZs and capture regions, most methods rely on online optimization that treats engagement zones and input bounds as constraints, leading to significant computational overhead and limiting their applicability in fast, decentralized, and resource-constrained environments, especially with multiple defenders. The differential game-based guidance strategies typically require the knowledge of the defender's policy and simplified models to obtain tractable feedback solutions without consideration of the attacker's physical input limits, which restricts their applicability when defenders act autonomously and their actuation saturates. Therefore, this work is motivated by the need to develop a tractable guidance law for the attacker that incorporates the geometry of EZs directly into the design and explicitly enforces the input constraints while guaranteeing safety and performance. To this end, the contributions of this work can be summarized as follows:
\begin{itemize}
    \item We propose a switching guidance strategy that prioritizes target interception when the attacker is outside all EZs and switches to an EZ avoidance mode when approaching its boundary.
    \item We propose a unified smooth safety constraint that aggregates multiple EZ threats via a log-sum-exp approximation, which provides a tunable parameter to prioritize the most critical zones.
    \item Our proposed guidance strategy incorporates input bounds into the design by augmenting the system with a smooth saturation function, guaranteeing that the attacker's lateral acceleration remains within its physical limits.
    \item We establish Lyapunov-based guarantees, proving forward invariance of the safe set together with convergence of the interception objective under the enforced input constraints, and validate the approach in multi-defender simulations.
\end{itemize}

\section{Problem Formulation}
\begin{figure}[h!]
    \centering
    \includegraphics[width=.5\linewidth]{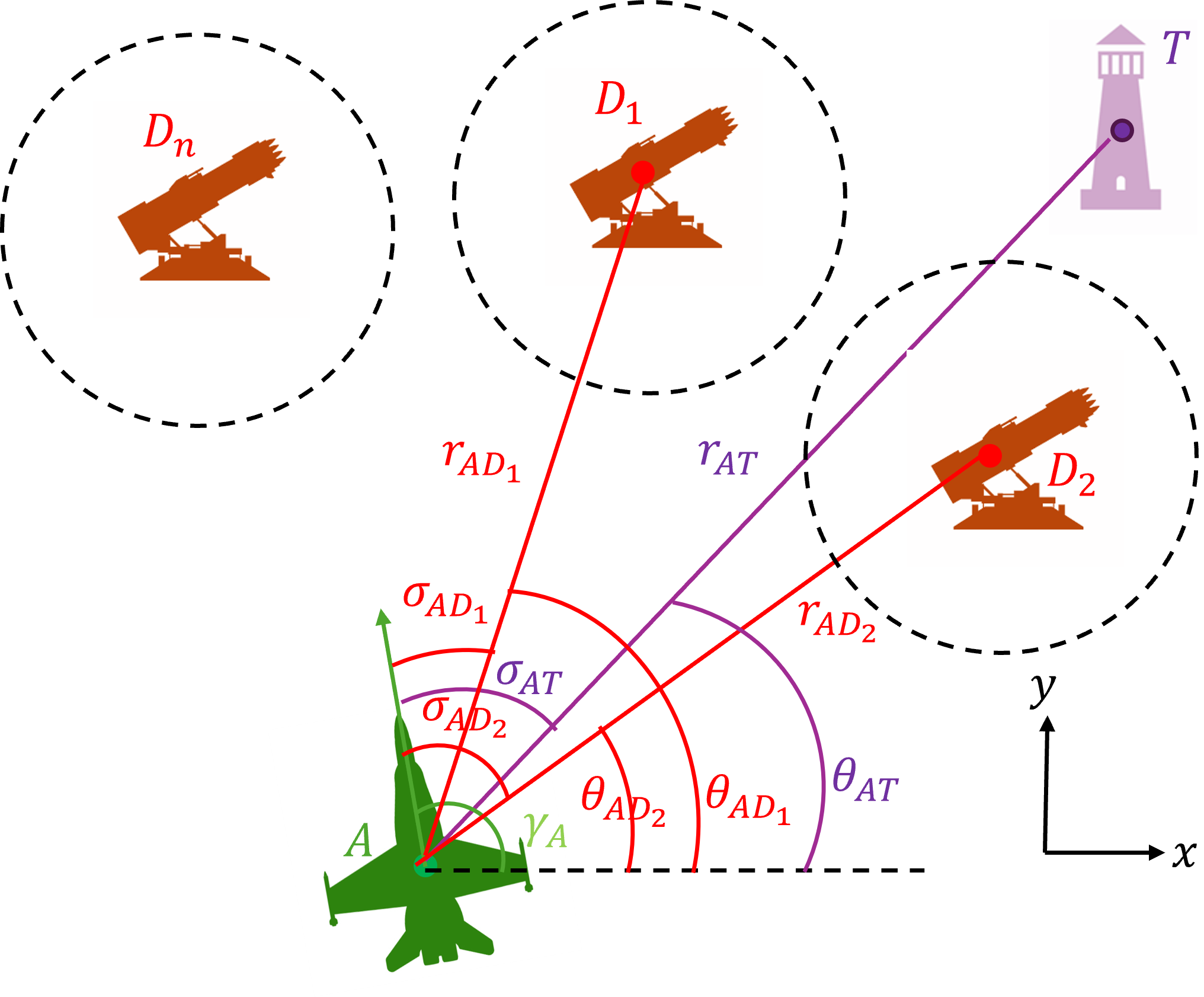}
    \caption{Attacker-Target-Defender engagement geometry.}
    \label{fig:enggeom}
\end{figure}

Consider a multi-agent system scenario consisting of an attacker $A$, $n$ defenders denoted by the set $\mathcal{D} = \{ D_1, D_2, \ldots, D_n \}$, and a stationary target $T$, as shown in \Cref{fig:enggeom}. The defenders are considered to be range limited, that is, able to travel a maximum distance of $R_j>0$ from their stationary point of origin $[x_{j}, y_{j}]^\top \in \mathbb{R}^2$ with a capture radius denoted by $c_j \in \mathbb{R}_{\geq 0}$ and have constant speeds $v_j> v_A \in \mathbb{R}_{>0}~$, $\forall~ j \in \mathcal{D}$. Such defenders may represent defense systems launched from stationary sites such as surface-to-air missiles (SAMs), naval interceptors, or other localized defense assets leading to constrained EZs.

Assuming the attacker to be a point mass non-holonomic vehicle moving with constant speed $v_A$, the equations of relative motion between the attacker and the $j$\textsuperscript{th} defender's point of origin or the target are governed by
\begin{subequations}
\begin{align}
    \dot{r}_{Aj} &=-v_A\cos\left(\gamma_A-\theta_{Aj}\right) ,\label{eqn:rel_dyn_1}\\
    r_{Aj}\dot{\theta}_{Aj} &= -v_A\sin\left(\gamma_A-\theta_{Aj}\right), \; \forall \; j \in\{\mathcal{D}, T\},\\
    \dot{\gamma}_A &= \frac{a_A}{v_A},
    \end{align}
    \label{eqn:rel_dyn}
\end{subequations}
where $r_{Aj}$ denotes the relative distance, $\theta_{Aj}$ denotes the line-of sight (LOS) angle from the attacker to the j\textsuperscript{th} entity, $\gamma_A \in(-\pi, \pi]$ denotes the attacker's heading angle and $a_A$ denotes its lateral acceleration, which is the control input for the attacker. Additionally, we define the attacker's lead angle as the angle subtended by the attacker's velocity to the respective LOS to defenders' origin or target, given as,
 \begin{align}
     \sigma_{Aj} = \gamma_A-\theta_{Aj},~ \forall \; j\in \{\mathcal{D}, T\}. \label{eqn:lead}
 \end{align}
In this paper, we aim to develop a guidance strategy that keeps the attacker safe from defenders by explicitly locating and avoiding parts of the state space where the attacker would be vulnerable to interception under its current control policy. These critical regions are referred to as EZs.
\begin{definition}[Engagement Zone (EZ)]\label{def:ez}
The EZ of a defender is defined as the set of the attacker's state-space from which the defender can guarantee capture, provided the attacker maintains its current heading and control strategy. 
\end{definition}
\begin{remark}
    Therefore, EZ characterizes the region in the attacker's configuration space where it is at risk of being intercepted by the defenders without altering its trajectory. In the relative polar coordinate frame centered at the point of origin of the defender, the \textit{engagement boundary} represents the outermost surface of the EZ, beyond which the attacker remains safe, but crossing which guarantees the interception of the attacker if it maintains its current trajectory. 
\end{remark}
For the fast defenders $D_i$, with speed ratio $\mu_i=\dfrac{v_A}{v_{D_i}} \in(0,1)$, the boundary of its EZ is characterized by,
\begin{align}
\rho_i = \mu_i R_i \left[ \cos \sigma_{A D_i} + \sqrt{\cos^2 \sigma_{A D_i} - 1 + \frac{(R_i + c_i)^2}{\mu_i^2 R_i^2}} \right], \label{eqn:eng_bound}
\end{align}
where $\rho_i$ denotes the critical radial distance from the defender $D_i$'s point of origin. Crossing this boundary places the attacker within the defender’s guaranteed capture region. Therefore, based on \Cref{def:ez}, the EZ of a defender $D_i$ with respect to the attacker is defined as, $\mathcal{E}_i := \left\{ (r_{AD_i}, \sigma_{AD_i}) \in \mathbb{R}_{\geq 0} \times (-\pi, \pi] \,\middle|\, r_{AD_i} \leq \rho_i(\sigma_{AD_i}) \right\}.$
% \begin{align}
% \mathcal{EZ}_i := \left\{ (r_{AP_i}, \sigma_{AD_i}) \in \mathbb{R}_{\geq 0} \times (-\pi, \pi] \,\middle|\, r_{AD_i} \leq \rho_i(\sigma_{AD_i}) \right\},
% \end{align}

The objective of this paper is to design a nonlinear guidance law for the attacker to intercept a stationary target while ensuring safety by avoiding all the defender-induced EZs and respecting the bounds on its lateral acceleration. Therefore, the attacker must satisfy three coupled objectives. First, the attacker needs to ensure \textit{target interception} by asymptotically driving its relative distance to the target to zero, that is, $\lim_{t \to \infty} r_{AT}(t) \to 0$. Second, the attacker must ensure \textit{EZ avoidance} by keeping its trajectory entirely outside each defender’s EZ at all times. Mathematically, this translates to the constraint, $(r_{AD_i}, \sigma_{AD_i}) \notin \mathcal{E}_i , \forall i \in \mathcal{D}, t\geq0$, which is similar to ensuring $r_{AD_i} > \rho_i$. Third, to ensure input feasibility, the guidance strategy must explicitly account for actuator limits by enforcing the following constraint on the attacker's lateral acceleration, $\vert a_A(t)\vert\leq  a_A^{\max}~ \forall t\geq0$, where $a_{\max}>0$ denotes the predefined symmetric constraints on its lateral acceleration.
\section{Main Results}
In this section, we formulate the safety conditions for the attacker in the presence of multiple defenders and derive the control law that guarantees both the target interception and avoidance of defender EZs under input constraints. To this end, we define a safety function for the attacker with respect to each EZ as $b_i= r_{AD_i}-\rho_i-\epsilon$, where $\epsilon\in\mathbb{R}_{>0}$ denotes a constant that represents the safety margin. Note that $b_i>0$ ensures the safety condition $r_{AD_i}>\rho_i$ with a clearance of at least $\epsilon$, while $b_i<-\epsilon$ implies that the attacker has penetrated inside the EZ and will be intercepted by the defender. Differentiating $b_i$ with respect to time and using \eqref{eqn:rel_dyn}, we obtain the dynamics of the safety function as
\begin{align}
\dot{b}_i &=-v_A\cos\sigma_{AD_i}-\frac{v_A}{r_{AD_i}}\nabla\rho_i\sin\sigma_{AD_i} - \nabla\rho_i\frac{a_A}{v_A} \label{eqn:bdot},
\end{align}
where $\nabla\rho_i$ denotes the gradient of engagement boundary of the $i$\textsuperscript{th} defender, computed using \eqref{eqn:eng_bound}, as
\begin{align}
\nabla \rho_i = -\mu_i R_i\sin\sigma_{AD_i} \left[1+\frac{\cos\sigma_{AD_i}}{\sqrt{\cos^2 \sigma_{A D_i} - 1 + \frac{(R_i + c_i)^2}{\mu_i^2 R_i^2}}}\right],    
\end{align}
To account for multiple EZs, we aggregate the safety function for all the EZs using the log-sum-exp operator 
\begin{align}
    h = -\beta\log\left(\sum_{i \in\mathcal{D}}e^{-b_i/\beta}\right)\label{eq:h}
\end{align}
where $\beta\in\mathbb{R}_{>0}$ is a user-defined constant that determines how close $h$ approximates the minimum values of individual safety functions.
\begin{remark}\label{rem:agg_safety_cond}
    The above safety aggregation function satisfies,
    \begin{align}
        \min_{i\in\mathcal{D}}b_i-\beta\log n \leq h \leq \min_{i\in\mathcal{D}}b_i,  \label{eqn:agg_safety_bounds}
    \end{align}
    where $n$ denotes the number of defenders. The upper bound implies that if $h>0$, then $\min_{i\in\mathcal{D}} b_i > 0$, which ensures $b_i > 0$ for all EZs (the attacker remains outside every defender’s EZ with the prescribed safety margin $\epsilon$). Conversely, if all $b_i>0$, then $h$ is guaranteed to be positive if, $\min_{i\in\mathcal{D}} b_i > \beta \ln n,$
% \begin{align}
%     \min_{i\in\mathcal{D}} b_i > \beta \ln n,
% \end{align}
which provides a design condition on the smoothing parameter $\beta$, obtained from the lower bound of $h$ in \eqref{eqn:agg_safety_bounds}. In particular, choosing $\beta$ sufficiently small makes $h$ a close approximation of $\min_i b_i$, thereby tightening the equivalence between the condition $h>0$ and the requirement $b_i>0~\forall~i \in \mathcal{D}$. On the other hand, selecting larger values of $\beta$ allows the aggregation to account for the combined effect of multiple EZs, rather than focusing solely on the nearest one.
\end{remark}
Differentiating \eqref{eq:h} with respect to time, we obtain the dynamics of the aggregate safety function as
% \begin{align}
%     \dot{h} =\sum_{i\in \mathcal{D}}w_i\dot{b}_i=\sum_{i\in \mathcal{D}}w_i\Big(&-v_A\cos\sigma_{AD_i}\nonumber\\&-\frac{v_A}{r_{AD_i}}\Delta\rho_i\sin\sigma_{AD_i} - \Delta\rho_i\frac{a_A}{v_A}\Big),
% \end{align}
\begin{align}
    \dot{h} =\sum_{i\in \mathcal{D}}\dfrac{e^{-b_i/\beta}}{\sum_{i\in\mathcal{D}}e^{-b_i/\beta}}\dot{b}_i = \sum_{i\in \mathcal{D}}w_i\dot{b}_i, \label{eqn:hdot}
\end{align}
such that $\sum_{i\in\mathcal{D}}w_i=1$ with $w_i\geq0$. 

We now propose the safety-preserving lateral acceleration command that ensures that the attacker does not enter the EZ boundaries as
\begin{align}
    a_A^b = \dfrac{v_AK_sh-v_A^2\displaystyle\sum_{i\in \mathcal{D}}w_i\Big(\cos\sigma_{AD_i}+\frac{\nabla\rho_i}{r_{AD_i}}\sin\sigma_{AD_i}\Big)}{\displaystyle\sum_{i\in\mathcal{D}}w_i\nabla\rho_i}, \label{eqn:safe_ctrl}
\end{align}
where $K_s\in\mathbb{R}_{> 0}$ denotes the controller gain for safety. For target interception, we utilize the attacker's command
\begin{align}
a_A^{T} = -K_I v_A \sigma_{AT} - \dfrac{v_A^2 \sin \sigma_{AT}}
{r_{AT}} \label{eqn:inter_law},
\end{align}
where $K_I\in\mathbb{R}_{> 0}$ is the controller gain for interception. These commands are obtained by setting $\dot{\sigma}_{AT}=-K_1\sigma_{AT}$ and $\dot{h}=-K_sh$, whose analysis is presented in the next theorem.

To handle the trade-off between safety and target-interception, we blend $a_A^b$ and $a_A^T$ via a switching function/convex combination of the form, 
\begin{align}
    a_A^d = \alpha(h) a_A^b + \left(1-\alpha(h)\right)a_A^T,
    \label{eqn:blend_ctrl}
\end{align}
to get the desired lateral acceleration for the attacker, where
\begin{align}
    \alpha(h) = \begin{cases}
1, & \text{if } h < \epsilon_h, \\
0, & \text{otherwise},
\end{cases}, \label{eqn:switch_fn}
\end{align}
with $\epsilon_h\in\mathbb{R}_{> 0}$ being a design parameter that defines the proximity threshold for safety. 

Under the combined law \eqref{eqn:blend_ctrl}, the safety term 
$a_A^b$ is activated in the region where $h$ is small to ensure that the attacker does not enter the engagement boundaries of the defenders, whereas outside that region ($h$ is large) the interception term $a_A^T$ dominates. To incorporate considerations of bounded control input during design, we introduce the symmetric version of the smooth input saturation model \cite{kumar2025provably}, as
\begin{align}
    \dot{a}_A = \left[1-\left(\frac{a_A}{a_{\max}}\right)^n\right]a_A^c - p_1 a_A, \label{eqn:sat_model}
\end{align}
where $n\geq2$, $p_1\in\mathbb{R}_{> 0}$ denote constants, and $a_A^c$ denotes the commanded lateral acceleration, which is the pseudo control input for the augmented system with the above saturation model. 
\begin{remark}\label{rmk:inputsat}
    From \eqref{eqn:sat_model}, it follows that when $|a_A|\to a_{\max}$, $\dot{a}_A\to -p_1a_A,$ and $|a_A|\to -a_{\max}$, $\dot{a}_A\to p_1a_A,$ which implies that $a_A$ will decrease if $a_A\to a_{\max}$ and $a_A$ will increase if $a_A\to -a_{\max}$. Therefore, the above saturation model ensures that $a_A$ remains bounded as $|a_A|<a_{\max}$. Thus, we augment the saturation model in \eqref{eqn:sat_model} with the kinematics \eqref{eqn:rel_dyn} and will design $a_A^c$, which will automatically ensure that $a_A$ remains within the permissible limits. 
\end{remark}
Let us define the error between the actual and the desired lateral acceleration as, $z =a_A-a_A^d$.
% \begin{align}
%     z =a_A-a_A^d \label{eqn:pseudi_ctrl}
% \end{align}
We now design the commanded lateral acceleration for the attacker as
\begin{align}
    a_A^c =& \dfrac{p_1a_A + \dot{a}_A^d+(\alpha-1)\frac{\sigma_{AT}}{v_A} +\alpha h\sum_{i\in \mathcal{D}}\frac{w_i\nabla\rho_i}{v_A} - K_a\mathrm{sign}(z)}{1-\left(\dfrac{a_A}{a_{\max}}\right)^n} \label{eqn:comd_acc}
\end{align}
where $K_a\in\mathbb{R}_{> 0}$ is the controller gain.

\begin{theorem}
    Consider the attacker-defender-target relative kinematics \eqref{eqn:rel_dyn}. If the attacker applies the commanded lateral acceleration  \eqref{eqn:comd_acc}, then it intercepts the target without entering the engagement boundaries of the defenders while respecting bounds on the control inputs.
\end{theorem}
\begin{proof}
    Consider the candidate for the Lyapunov function $ V_1=\frac{1}{2}\left(1-\alpha(h)\right)\sigma_{AT}^2 + \frac{1}{2}\alpha(h)h^2 .$
    % \begin{align}
    % \dot{V}_1 &= \frac{1}{2}\sigma_{AT}^2\alpha\left(1-\alpha\right)\dot{h} + \left(1-\alpha\right)\sigma_{AT}\dot{\sigma}_{AT} \nonumber \\&-\frac{1}{2} \alpha\left(1-\alpha\right) \dot{h}h^2 + \alpha h\dot{h}  \nonumber \\&=\frac{\dot{h}\alpha\left(1-\alpha\right)}{2}\left[\sigma_{AT}^2-h^2\right] \nonumber\\&+ \left(1-\alpha\right)\sigma_{AT}\left(\frac{\alpha(h) a_A^b + \left(1-\alpha(h)\right)a_A^T}{v_A}+\frac{v_A}{r_{Ai}}\sin\sigma_{Ai}\right)\nonumber \\ &+\alpha h\left(\sum_{i\in \mathcal{D}}w_i\left[-v_A\cos\sigma_{AD_i}-\frac{v_A}{r_{AD_i}}\Delta\rho_i\sin\sigma_{AD_i} - \Delta\rho_i\frac{a_A}{v_A}\right]\right)
    % \end{align}
    Differentiating $V_1$ with respect to time and using \eqref{eqn:bdot} and \eqref{eqn:hdot}, we obtain,
    \begin{align}
        \dot{V}_1 =& (1-\alpha)\sigma_{AT}\dot{\sigma}_{AT} + \alpha h\dot{h}  \nonumber \\
        % &=\left(1-\alpha\right)\sigma_{AT}\left(\frac{a_A}{v_A}+\frac{v_A}{r_{AT}}\sin\sigma_{AT}\right)\nonumber \\ &-\alpha h\sum_{i\in \mathcal{D}}w_i\Bigg[v_A\cos\sigma_{AD_i}+\frac{v_A\Delta\rho_i}{r_{AD_i}}\sin\sigma_{AD_i} + \frac{a_A\Delta\rho_i}{v_A}\Bigg]\nonumber \\
        =&\left(1-\alpha\right)\sigma_{AT}\left(\frac{a_A^d}{v_A}+\frac{v_A}{r_{AT}}\sin\sigma_{AT}\right)\nonumber \\ &-\alpha hv_A\sum_{i\in \mathcal{D}}w_i\Bigg[\cos\sigma_{AD_i}+\frac{\nabla\rho_i}{r_{AD_i}}\sin\sigma_{AD_i} + \frac{a_A^d\nabla\rho_i}{v_A^2}\Bigg]\nonumber \\ &+\left((1-\alpha)\frac{\sigma_{AT}}{v_A} -\alpha h\sum_{i\in \mathcal{D}}\frac{w_i\nabla\rho_i}{v_A}\right) z
    \end{align}
    using $a_A=z + a_A^d$. On substituting the desired lateral acceleration \eqref{eqn:blend_ctrl} in the above equation, we obtain 
    \begin{align}
        \dot{V}_1 =& -(1-\alpha)^2K_1\sigma_{AT}^2 -\alpha^2K_sh^2 \nonumber\\&+ \alpha(1-\alpha)\Bigg[\sigma_{AT}\left(a_A^b+\frac{v_A}{r_{AT}}\sin\sigma_{AT}\right)\nonumber\\&-v_Ah\sum_{i\in\mathcal{D}}w_i\Bigg(\cos\sigma_{AD_i}+\frac{\Delta\rho_i}{r_{AD_i}}\sin\sigma_{AD_i} + \frac{a_A^T\Delta\rho_i}{v_A^2}\Bigg) \Bigg]\nonumber \\ &+\left((1-\alpha)\frac{\sigma_{AT}}{v_A} -\alpha h\sum_{i\in \mathcal{D}}\frac{w_i\nabla\rho_i}{v_A}\right) z,
    \end{align}
    which simplifies to $\dot{V}_1 = -(1-\alpha)^2K_1\sigma_{AT}^2 -\alpha^2K_sh^2 +\left((1-\alpha)\frac{\sigma_{AT}}{v_A} -\alpha h\sum_{i\in \mathcal{D}}\frac{w_i\nabla\rho_i}{v_A}\right) z$,
    % \begin{align}
    %     \dot{V}_1 &= -(1-\alpha)^2K_1\sigma_{AT}^2 -\alpha^2K_sh^2 \nonumber\\&+\left((1-\alpha)\frac{\sigma_{AT}}{v_A} -\alpha h\sum_{i\in \mathcal{D}}\frac{w_i\nabla\rho_i}{v_A}\right) z
    % \end{align}
    since $\alpha+(1-\alpha)=0$.

    Since $\sigma_{AT}\to 0$, $\dot{r}_{AT}\to -v_A$, which eventually leads to target interception. Similarly, when $h\to 0$, the RHS of \eqref{eq:h} tends to $0$, which is possible only when $b_i\to 0$ because $\beta\in\mathbb{R}_{>0}$. This essentially means that the attacker is moving away from an EZ. Therefore, the proposed strategy simultaneously guarantees target interception under safety constraints.
    
    To incorporate the input constraint, we consider another Lyapunov candidate function, $V=V_1+V_2$, where $V_2 =z^2/2$. On differentiating $V_2$ with respect to time and using \eqref{eqn:sat_model}, 
    \begin{align}
        \dot{V}_2=& \dot{V}_1 + \dot{V}_2 = \dot{V_1}+z\dot{z}=\dot{V_1}+z\left(\dot{a}_A-\dot{a}_A^d\right) \nonumber\\ =&
       -(1-\alpha)^2K_1\sigma_{AT}^2 -\alpha^2K_sh^2 +\Bigg\{(1-\alpha)\frac{\sigma_{AT}}{v_A} \nonumber\\&-\alpha h\sum_{i\in \mathcal{D}}\frac{w_i\nabla\rho_i}{v_A}+\left[1-\left(\frac{a_A}{a_{\max}}\right)^n\right]a_A^c - p_1 a_A -\dot{a}_A^d\Bigg\} z.
    \end{align}
    Choosing the commanded acceleration proposed in \eqref{eqn:comd_acc} renders the derivative of the Lyapunov function candidate,
    \begin{align}
        \dot{V}_2 = -(1-\alpha)^2K_1\sigma_{AT}^2 -\alpha^2K_sh^2 - K_az^2 .\label{eqn:finalv_dot}
    \end{align}
    It follows from the above equation that when the attacker is away from all the EZs ($\alpha=0$), $\dot{V}_2=-K_1\sigma_{AT}^2-K_a z^2$. If $K_T$ and $K_a$ are chosen positive, then $\dot{V}_2$ becomes negative definite. This implies $\sigma_{AT}\to 0$ and $z\to 0$ asymptotically, resulting in $\dot{r}_{AT}=-v_A<0$ from \eqref{eqn:rel_dyn_1} leading the attacker to intercept the target. When the attacker is near any of the EZs ($\alpha=1$), then \eqref{eqn:finalv_dot} yields, $\dot{V}_2=-K_sh^2-K_az^2$, which is negative definite in $(h,z)$ for $K_a, K_s>0$. This ensures $h\to 0$ and $z\to 0$ as $t \to \infty$. Particularly, the term $-K_sh^2$ ensures $\dot{h}\geq-K_sh$ so that $h(t)$ decays exponentially to zero while remaining non-negative. Hence, the safe set $\{h\geq0\}$ is forward invariant, which implies that if $h(0)>0$ then $h(t)>0, \forall\; t \geq0$. This implies that the attacker never enters any of the EZs, since $h>0$ implies $\min_{i \in \mathcal{D}}b_i>0$ from \Cref{rem:agg_safety_cond}. Further, inferences from \Cref{rmk:inputsat} show that the actual control $a_A$ never exceeds the permissible limits. %This concludes the proof.
\end{proof}

\begin{remark}
To ensure smooth transitions and avoid potential instability due to discontinuous switching during implementation, we employ a continuous switching function given by, $\alpha(h) = \frac{1}{1+ e^{\frac{h-\epsilon_\alpha}{\delta}}}$, %     \begin{align}
%     \alpha(h) = \frac{1}{1+ e^{\frac{h-\epsilon}{\delta}}}
% \end{align}
where $\epsilon_\alpha$ and $\delta$ are positive constants. For the above switching function, it can be observed that $\alpha \in (0, 1)$, with $\alpha(\epsilon_\alpha)=1/2$, $\alpha \approx 1$ when $h\ll\epsilon_\alpha$ and $\alpha \approx 0$ when $h\gg\epsilon_\alpha$. 
\end{remark}
This smooth formulation ensures a gradual transition between safety and interception actions, thereby avoiding abrupt control changes. In particular, it guarantees that the derivative term $\dot{a}_A^d$ in the commanded input \eqref{eqn:comd_acc} remains bounded, which is essential for practical realizability. For analytical simplicity, the stability proof above was presented using the discontinuous switching law. A complete proof under the smooth switching formulation will be included in our extended version. Nonetheless, the existing proof remains sufficient because the smooth case can be interpreted as a composition of three regimes: (i) the region where 
$a_A^b$ is fully active, (ii) the region where $a_A^T$ is fully active, and (iii) the transition region where a convex combination of both is applied. Since the first two regimes coincide with the discontinuous proof and the transition region preserves convexity of the Lyapunov arguments, the stability and safety guarantees continue to hold.

\section{Simulation Results}
\begin{figure*}[h!]
    \centering
    \begin{subfigure}[t]{.245\linewidth}
    \centering
    \includegraphics[width=\linewidth]{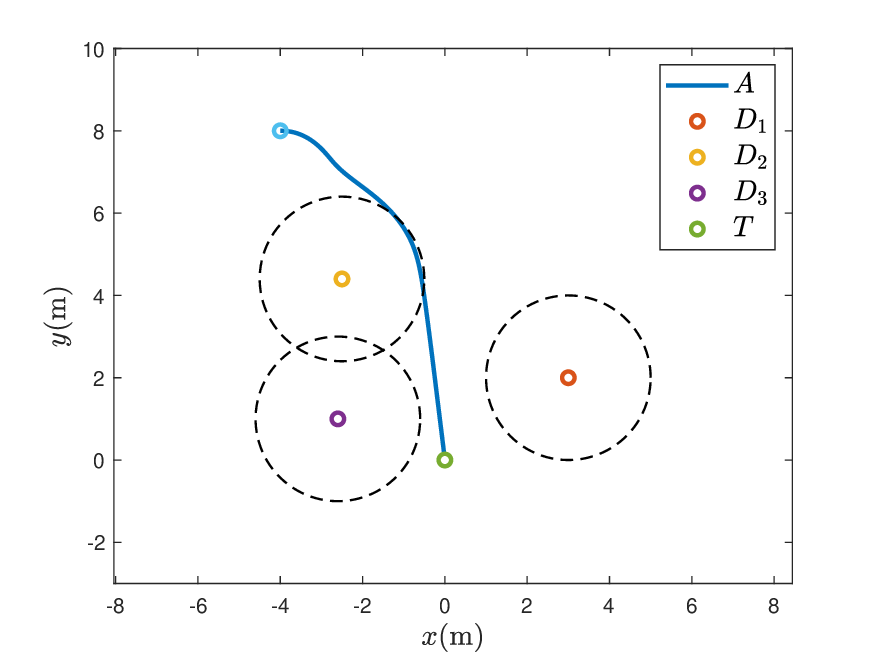}
    \caption{Trajectories.}
    \label{fig:traj_ini}
    \end{subfigure}
    \begin{subfigure}[t]{.245\linewidth}
    \centering
    \includegraphics[width=\linewidth]{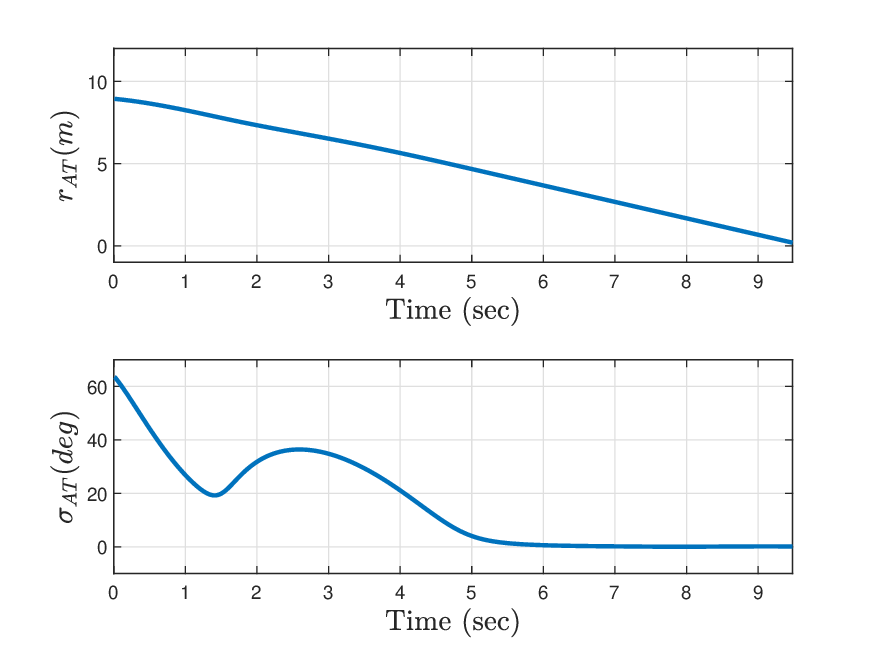}
    \caption{Range and bearing angle.}
    \label{fig:var_ini}
    \end{subfigure}
    \begin{subfigure}[t]{.245\linewidth}
    \centering
    \includegraphics[width=\linewidth]{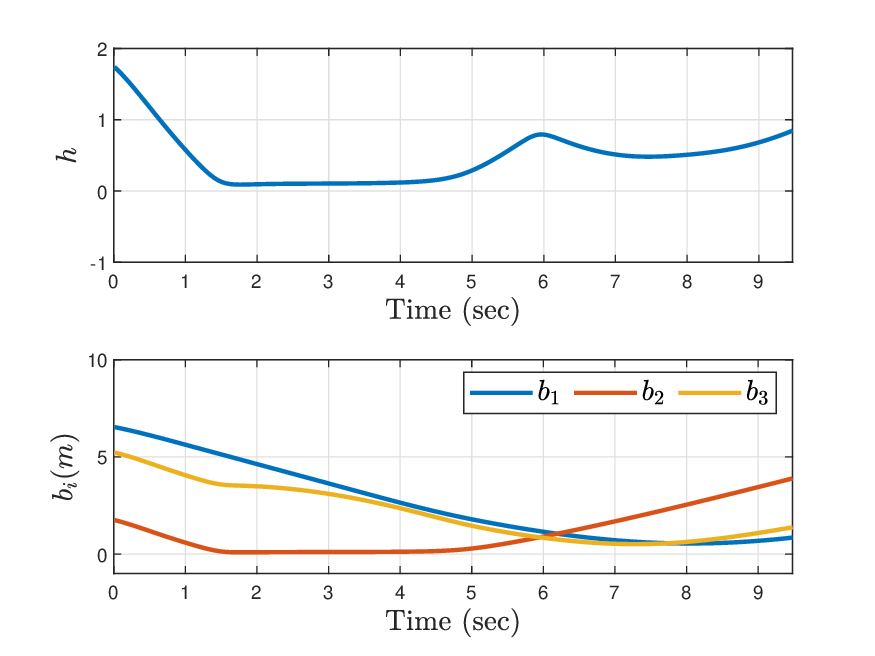}
    \caption{Safety constraints.}
    \label{fig:saf_ini}
    \end{subfigure}
    \begin{subfigure}[t]{.245\linewidth}
    \centering
    \includegraphics[width=\linewidth]{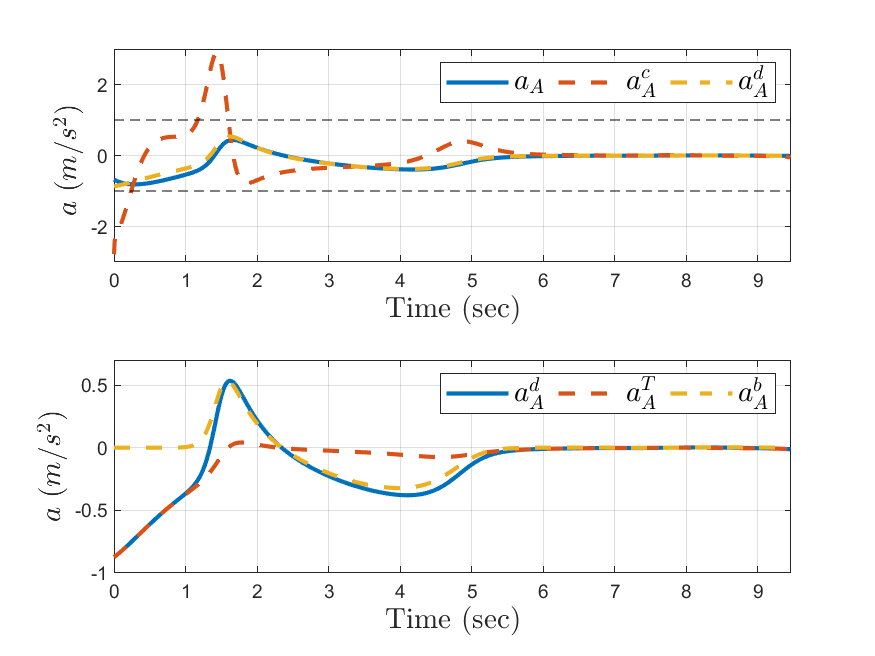}
    \caption{Attacker's lateral acceleration.}
    \label{fig:ctrl_ini}
    \end{subfigure}
    \caption{Target interception in the presence of three defenders ($[x_A(0), y_A(0)]^\top = [-4, 8]^\top$ m).}
    \label{fig:ini_1}
\end{figure*}

\begin{figure*}[h!]
    \centering
    \begin{subfigure}[t]{.245\linewidth}
    \centering
    \includegraphics[width=\linewidth]{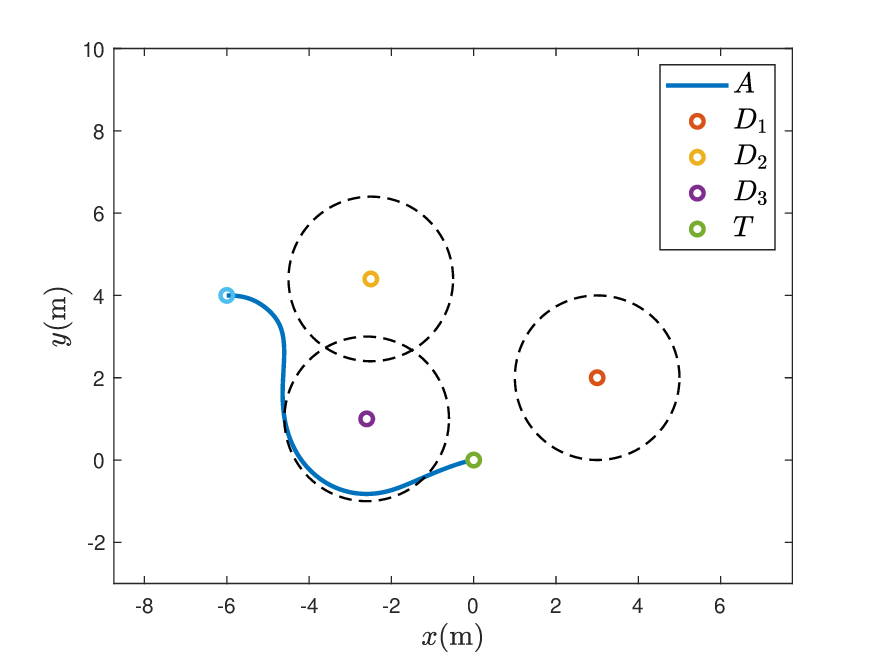}
    \caption{Trajectories.}
    \label{fig:traj_ini_2}
    \end{subfigure}
    \begin{subfigure}[t]{.245\linewidth}
    \centering
    \includegraphics[width=\linewidth]{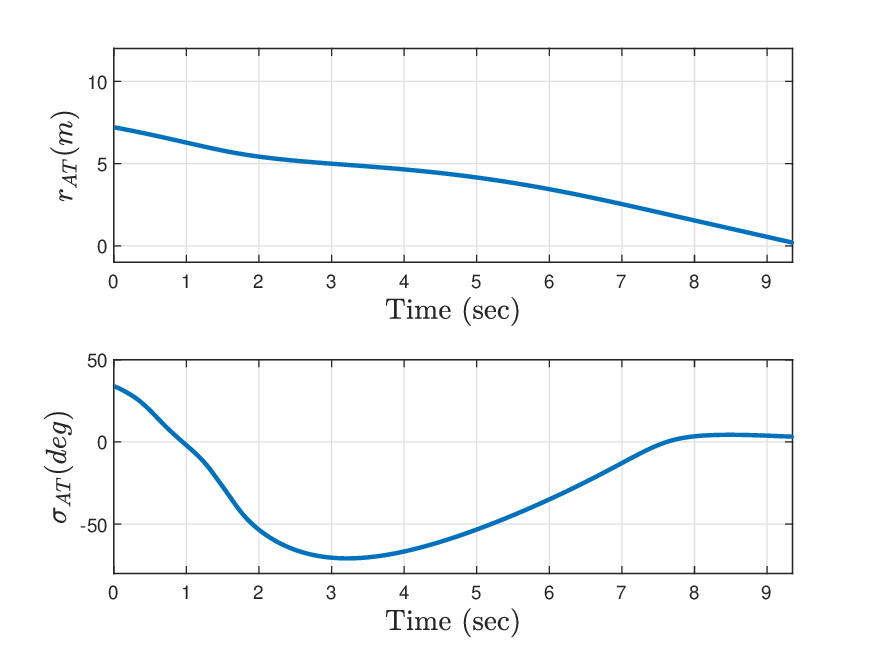}
    \caption{Range and bearing angle.}
    \label{fig:var_ini_2}
    \end{subfigure}
    \begin{subfigure}[t]{.245\linewidth}
    \centering
    \includegraphics[width=\linewidth]{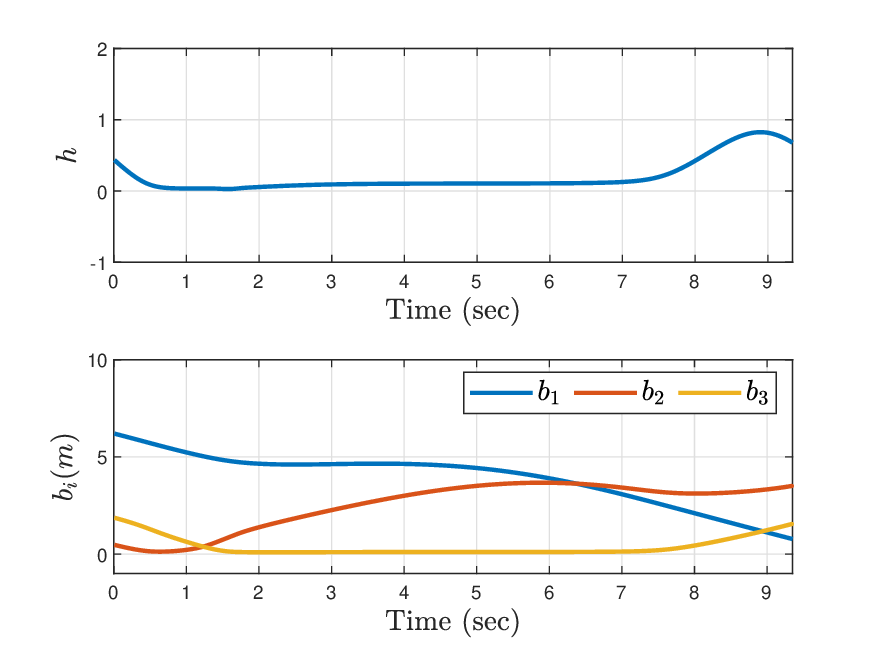}
    \caption{Safety constraints.}
    \label{fig:saf_ini_2}
    \end{subfigure}
    \begin{subfigure}[t]{.245\linewidth}
    \centering
    \includegraphics[width=\linewidth]{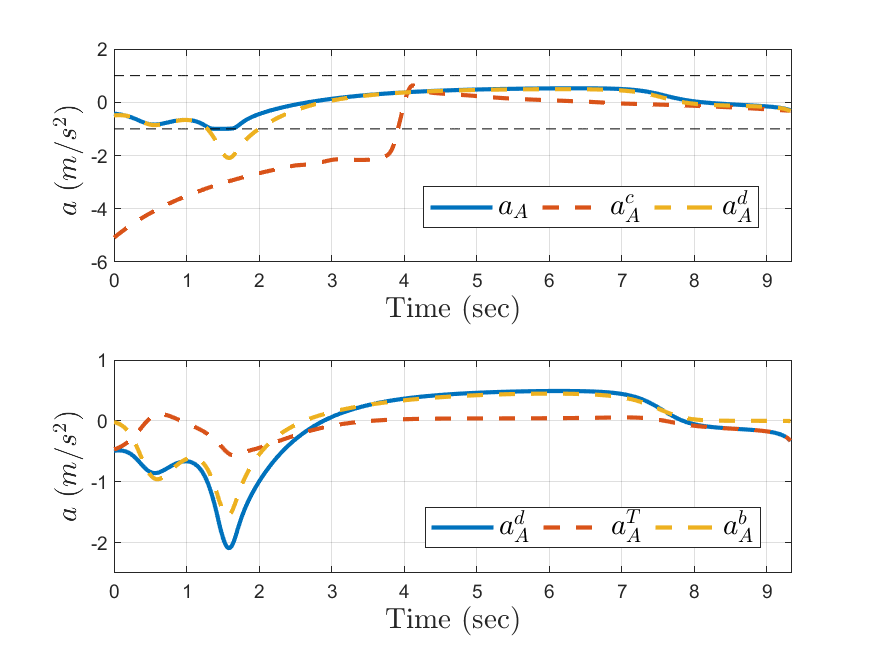}
    \caption{Attacker's lateral acceleration.}
    \label{fig:ctrl_ini_2}
    \end{subfigure}
    \caption{Target interception in the presence of three defenders ($[x_A(0), y_A(0)]^\top = [-6, 4]^\top$ m).}
    \label{fig:ini_2}
\end{figure*}

In this section, we present the simulation results to demonstrate the efficacy of our approach in intercepting a stationary target in the presence of three defenders. The origin of the defenders are located at \([x_{D_1}, y_{D_1}]^\top = [3,\, 2]^\top\), \([x_{D_2}, y_{D_2}]^\top = [-2,\, 4.4]^\top\), and \([x_{D_3}, y_{D_3}]^\top = [-2.6,\, 1]^\top\) m, respectively. All defenders are assumed to be homogeneous in their capabilities, with a speed ratio of $\mu_i = 0.7$ relative to the attacker, have a maximum engagement range $R_i = 1.5\,\text{m}$, and a capture radius $c_i = 0.5\,\text{m}$. Therefore, the total maximum range of each defender is  $R_i + c_i = 2\,\text{m}$ from their point of origin as depicted by dotted circles in the trajectory plots in \Cref{fig:traj_ini,fig:traj_ini_2}. In all the simulations, the attacker moves at a constant speed of $1$ m/s. The attacker's controller gains are selected as $K_s=0.3, K_T=0.7$ and $K_a=3.5$, while the other design parameters are chosen as $\beta=0.3, \epsilon_\alpha=0.1, \delta=0.1$. The parameters of the input saturation model in \eqref{eqn:sat_model} are selected as $p_1=0.2$, $n=2$, with lateral acceleration bounded as $\vert a_A\vert<a_A^{\max}=1$ m/s\textsuperscript{2}. 

In the first set of simulation results, the attacker starts at $[x_A, y_A]^\top = [-4, 8]^\top$ m, such that the initial LOS from the attacker to the target intersects only with the maximum engagement range of defender $D_2$. The resulting attacker's trajectory is shown in \Cref{fig:traj_ini}, where it briefly crosses the boundary of the defender's maximum range before successfully intercepting the target. The aggregated and individual safety constraints for the pursuer with respect to each defending zone are depicted in \Cref{fig:saf_ini}, where $h>0$ and $b_i>0,\; \forall i\in\mathcal{D}$. This confirms that even when the pursuer enters or crosses the defender's range, it never reaches a state where its interception by the defender is possible. Therefore, compared to previous approaches that enforce conservative safety margins preventing entry into the defender’s range, the proposed strategy enables faster target interception by allowing controlled penetration of the EZ without compromising safety. 

The attacker-target relative range and bearing angles are depicted in \Cref{fig:var_ini}, which at steady state converge to zero, demonstrating successful interception of the target. During the transient phase, deviations from convergence to zero arise when the attacker traverses the defender's engagement range (for e.g., see $\sigma_{AT}$ from $t=1.5$ s to $t=5.2$ s), reflecting the safety-preserving maneuvers executed by the attacker based on our proposed strategy. \Cref{fig:ctrl_ini} compares the control inputs of the attacker, where the first subplot compares actual, desired, and commanded accelerations, which converge to the same value at the steady state, resulting in $z \to 0$ consistent with our theoretical results. Additionally, while the commanded and desired acceleration may exceed the allowable limits (as depicted by the black dotted lines), the actual lateral acceleration remains within the physical bounds due to the saturation model incorporated into the design. The second subplot in \Cref{fig:ctrl_ini} depicts the target interception and safety components of the attacker's desired lateral acceleration, where the dominant term shifts with the defender threat (see, for e.g., from $t=1.5-5.2$ s, $a_A^b$ dominates to ensure safety, while at other times $a_A^T$ dominates to achieve interception.) 

In the second set of results, as in \Cref{fig:ini_2}, the attacker starts at 
$[x_A, y_A]^\top = [-6, 4]^\top$ m, such that the attacker-target LOS intersects with two overlapping defenders' ranges. Unlike the first case, where the attacker primarily encountered a single EZ at a time, this scenario presents a concave notch where the influence of two defenders strongly affects the attacker’s motion. As shown in \Cref{fig:traj_ini}, the attacker safely maneuvers around this region and achieves successful interception of the target, highlighting the capability of the proposed strategy to handle multiple EZs simultaneously. The profile of the attacker's actual lateral acceleration is illustrated in \Cref{fig:ctrl_ini_2}, which remains within the allowable bounds for all time. However, a brief saturation in the lateral acceleration is observed between 1 and 2 seconds due to the high acceleration demand required to maneuver away from the concave notch between the overlapping zones.

\section{Conclusions}
We designed a nonlinear guidance law for an attacker to safely intercept a target while avoiding capture by the defenders and respecting the physical bounds on the attacker's lateral acceleration. The defender-induced regions where attacker interception by defenders is guaranteed were modeled as Engagement Zones (EZs) and directly incorporated into the guidance design, along with a smooth saturation model to ensure control input feasibility. A smooth minimum (log-sum-exp) function was adopted to aggregate risks from multiple zones in a unified safety measure. Lyapunov analysis under the idealized discontinuous switching function established safe-set invariance near EZs and asymptotic target interception when away from the EZs, all while respecting input bounds. Numerical simulations with multiple defenders validated the approach across diverse initial conditions, including challenging configurations with overlapping zones and concave notches. Our future extended version will include the stability analysis to fully discontinuous switching, investigate gain and parameter selection for performance vs conservativeness trade-offs, and explore moving targets/defenders' point of origin, 3D engagements, and hardware experiments for real-time validation.

\bibliographystyle{ieeetr}
\bibliography{references}

\end{document}